# Positron Annihilation Study of Biopolymer Inulin for Understanding its Structural Organization

Bichitra Nandi Ganguly,*[1] Madhusudan Roy, and S.P. Moulik[2]

[1]*Saha Institute of Nuclear Physics, 1/AF Bidhannager, KOLKATA, INDIA.*

[2]*Physical Chemistry Department , Jadavpur University KOLKATA, INDIA.*

**Corresponding author : Bichitra Nandi Ganguly*

*Fax No: +91 33 23374637, email : bichitra.ganguly@saha.ac.in*

## ABSTRACT

Inulins are nano-meter size semi-crystalline particles, composed of oligomeric fructose units. It has been subjected to fine micro-structural analysis under temperature variations using mainly positron annihilation spectroscopy. The results show a non-monotonous temperature sensitive behaviour of the positron parameters, with considerable variation of its free volume size. The ortho-positronium pick-off component shows a major thermotropic transition at ~320K and a structure loss due to glass transition.

Differential scanning calorimetry confirms the onset of the major molecular transition around the same temperature with an enthalpy change of ΔH ~379J /gm and thermo-gravimetric analysis shows mass loss in the said transition.

## 1. INTRODUCTION:

*Inulins* are a group of naturally occurring *polysaccharides* produced by plants [1], composed mainly of fructose units, and typically have a terminal glucose unit. The fructose units in this compound are joined by a β(2→1) glycosidic bond, as shown in scheme 1. It is valued because of its medicinal importance and often used as dietary fibre in the food products [2-4]. It is considered suitable for diabetics and is potentially helpful in managing blood sugar-related illnesses. Also, one of the important uses of inulin and its analogues is to help measure kidney function by determining the glomerular filtration rate (GFR), in clinical studies [5]. It may be pertinent to mention that inulin is generally recognized as safe (GRAS) by the U.S. Food and Drug Administration (FDA) [6]. As a means of non conventional energy resources, non hydrolyzed inulin can also be directly converted to ethanol in a simultaneous saccharification and fermentation process, which may have great potential for converting crops high in inulin into ethanol for fuel[7] .

Having such multifarious applications, an extensive physico-chemical study of such an important polymeric ingredient in liquid dispersive phase has been reported only recently [8]. Some of the solid state physical and water absorption properties along with X-ray diffraction studies on inulin have been performed earlier and a detailed report[9] studying the crystalline to semi crystalline form for applications deserves mention. An extensive review [10] regarding the structural aspects revealed orthorhombic and pseudo-hexagonal geometry for hydrated and semi- hydrated (only half a molecule of water per fructosyl unit) inulin respectively, consisting of two anti parallel six fold helices. Since the molecular chain structure is lined up through oxygen bridges (as in figure 1.) and shows a stacking pattern, it is of interest to investigate the temperature dependent structural parameters of the bio-polymer inulin. The physical properties during storage is of crucial importance for the shelf-life of food and pharmaceutical products containing the biopolymer in the "dry" state. To meet such requirements and to analyse the stable configuration for storage at varying temperatures in "dry"(may be with the inherent semi - hydrated )conditions [11], it necessitates for a detailed micro- structural study at molecular level using a suitable microprobe.

Perhaps it becomes worthwhile to have a quick assay of the crystalline structure and molecular units for each variety of samples (since a lot of variation of Inulin sample exists and storing conditions also vary as quoted in the literature), before a detailed structure analysis is made. For this reason, the as received inulin sample (stored under vacuum desiccation) has been first examined through X-ray diffraction method and FT- IR spectroscopy for its functional units.

Further, since positron annihilation is a viable and well established technique to study electronic micro-environment (due to changes in electron density), as is known for its applicability towards microstructure analysis of polymers [12] especially for their glass transition temperatures [13], it has been used here to elucidate the temperature dependent molecular micro-structural properties of inulin. Study of electron momentum distribution [14] through Doppler broadening of positron annihilation radiation energy (line shape parameters 'S' and 'W') over the temperature domain 273 to 360 K give an alternative assay of the molecular configuration as well. Important features like changes in free volume in the gradual heating cycle have also been observed through positron annihilation life time, such detailed studies in of inulin, has not been reported before.

Additionally, temperature dependent complementary techniques like thermal analysis by Differential Scanning Calorimetry (DSC) and Thermo-gravimetry (TG) have been used to confirm and to understand the configuration changes in the system, separately.

## 2. EXPERIMENTAL METHODS *:*

*a) Material Characteristics:*

A pure variety of Inulin from chicory has been procured from Sigma life science (USA), Product Number : I2255 , which is hygroscopic and therefore was stored under vacuum desiccation. The molecular structural formula is given in Scheme 1. In a previous study [8] the authors determjined the molar mass of the inulin to be 4458; the3 degree of polymerization (n) was thus 25. It has been constantly checked for its moisture content, by simple and accurate weighing method for a test fraction of sample before experimentation. No weight loss/gain has been observed at ordinary conditions (while storing at room temperature 298 K, in vacuum); the sample was semi-hydrated by its nature as stated [10].

*b) X-ray Diffraction measurements:*

The inulin powder sample was subjected to X-ray diffraction analysis using Rigaku diffractometer with CuKα radiation (wavelength of the radiation, k = 1.54 Å). The data have been collected in the range (2θ)~ $5^o$ –$60^o$ with a step size of $0.02^o$ at room temperature. Si has been used as external standard. The XRD pattern has been shown in **Figure 2**. The grain size of the powder sample has been calculated using Scherrer formula [15 ]: $D_{hkl} = K\lambda/\beta_P \cos\theta$ , where, $D_{hkl}$ is the average grain size, K the shape factor (taken as 0.9), λ is the X-ray wavelength, $\beta_P$ is the full width at half maximum (FWHM) intensity after correction of instrumental resolution and θ is the Bragg angle chosen at ~$22^o$.

*c) Fourier transmission infrared (FT-IR) spectra:*

FT-IR spectra of the Inulin powder [16] (as pellets in KBr, without moisture) were recorded using a Fourier transform infrared spectrometer (Perkin Elmer FTIR system; Spectrum GX) in the range of 400-6000 $cm^{-1}$ with a resolution of 0.2 $cm^{-1}$ as shown in **Figure 3**.

*d) Positron annihilation spectroscopy (PAL):*

**Lifetime measurements** : A fast–fast gamma-gamma coincidence circuit was used for the purpose of measuring positron annihilation lifetime spectrum using spectroscopic quality $BaF_2$ as the scintillators (cone-shaped of dimension 25.4mm diameter tapered to 12.7mm diameter and 25.4mm height ) coupled to XP2020Q photomultiplier tubes as the detectors, with the carrier free positron source ($Na^{22}$ in the form of NaCl, ~5 μ Ci point source, on an ultra thin Ni foil~3μm and folded) and placed inside the packed samples in a degassed sample chamber under vacuum with different temperature settings using temperature controlled thermostat (±0.1K). A total of $10^6$ counts were recorded under the spectrum with a peak to back ground ratio 5000:1. The resolution of the spectrometer was ~260 ps, as measured by $Co^{60}$ source at the positron window settings.
The spectrum was analyzed by PATFIT program [17], with a necessary source correction ~5 % to get the final analysis of the data. Mainly three life time components were found in all the cases, with a good variance of fit ~1; the shortest life time $\tau_1$ with its intensity $\mathbf{I_1}$ varying at different temperature is ascribed to annihilation in flight along with short lived para-positronium component. The second or the middle life time components $\tau_2$ with corresponding intensities $\mathbf{I_2}$ at different temperatures are given in **Figure 4 a and b**. **Figure 5**, reveals that there is a significant fraction of *ortho*-positronium pick-off life time component, greater than 1ns (**Figure 5a**), at each temperature regime and the corresponding intensities have been mapped in **Figure 6,** showing also the average long lived component, the discussions of which are relegated under the following section **3.**

**Doppler broadening (DB) spectroscopy**: The energy distribution of the electron-positron annihilation gamma radiation near 0.511 MeV photon is measured in this method. The detector used in DB spectrometer is a close ended co-axial high purity Ge (HPGe) detector (p-type) from ORTEC. The germanium crystal having an active volume of 70cc has been kept at liquid nitrogen temperature and the detector was well shielded in order to minimize the room background effect. The detector has been biased at 2000V positive potential. The signal from the detector via preamplifier passes to the spectroscopic amplifier and then to the MCA. The channel calibration of MCA is done with the help of standard sources namely $^{60}Co$ with a gamma photopeak at 1.33 MeV, the energy resolution (FWHM ~1.75 keV), and with $^{207}Bi$ for the gamma photopeak at 579 keV with the energy resolution (FWHM~ 1.1keV). The experimental spectrum was recorded around 511 keV annihilation line in order to determine the 'S' parameter, this being evaluated as the ratio of counts in the central region (511±1.7 keV) of the spectrum to the total counts recorded under the spectrum. Also 'W' parameter was measured at the wings (2keV< |E-511| <7keV) for the different samples on either side of the Doppler spectrum to the total fixed counts under the spectrum. The results are shown in **Figure 7a and b**.

*e) Themal analysis: Differential Scanning Calorimetry(DSC)and Themogravimetry(TG):*

Calorimetric measurements has been performed using a differential scanning calorimeter (DSC: Model No. 204 F1, NETZSCH, Germany) in the range of 270 – 440 K, purging with 99.999 % pure nitrogen gas. **Figure 8a** represents the DSC scan of the sample in the range of 270 - 440 K initiated with heating, followed by cooling, with a rate of 5K $min^{-1}$

The TG measurements have been taken using a thermo-gravimetric analyzer (TGA) of Netzsch, Germany (model: STA 449C) using 99.999% percent pure nitrogen as purge. The analysed results are given in **Figure 8b**.

## 3. RESULTS AND DISCUSSION:

Initial characterization of the Inulin material as received, has been subjected to the methods of analysis like X-ray diffraction and FT –IR Spectroscopy for its structural assay, before it was subjected to finer analysis such as positron annihilation spectroscopy. The crystallinity of Inulin sample has been detected and the characteristic peak for the Bragg angle around 22° (shown in **Figure 2**.) was used for the particle size determination by Debye –Scherrer method[15] and was found to be ~8 nm. Partial crystallinity could be present in the sample, the estimated degree of crystallinity in the present case was found be ~ 28.22 %. The partial crystallinity in semi hydrated Inulin sample was also reported earlier [8] with hexagonal characteristics, but orthorhombic and pseudo-hexagonal geometry was reported by others[10]. It was also noticed [18] that the difference between the hydrated and the semi-hydrated unit cells does not seem to correspond to any change in the conformation of inulin, but rather to a variation of water content.

However, stability, variation of the crystallinity versus water content and glass transition temperature was studied [9] for its usage as a food ingredient.

For elucidation of the main functional units of the Inulin compound as received, the characteristic infra-red spectrum was recorded (shown in **Figure 3**.). The polymeric chain of Inulin consists principally of linear chains of fructosyl units linked by a $\beta$ (2→1) bonds and terminal ends by a glycosyl unit. In the **Figure 3,** the carbohydrate region 900-1200 shows overlapping broad absorption band with a sharp maximum around 1030cm$^{-1}$. In 2700 to 4000 cm$^{-1}$ region of the spectrum, the terminal –OH groups in the molecular structure [16] along with fructose units show the strong stretching frequencies around 3600 cm$^{-1}$, with a shoulder at 2930 cm$^{-1}$.

**a) Positron Annihilation Life time (PAL) study**:

For the detailed microstructure of the *polysaccharide* supra-structure analysis, the inulin sample was subjected to temperature dependent positron annihilation spectroscopic study. The temperature variation of the experimental life time spectrum in each of the cases could be resolved in to three components. Of the two positron components, the $\tau_1$ is the shortest(fastest) component, around 200ps could be assigned for free annihilation or with a part of *para*-Positronium fraction with an intensity $\mathbf{I}_1$ ranging from (30-40% fraction on the average) and are not important in the context of molecular structure, but the middle component $\tau_2$ shows a periodic pattern along with it its intensity represented by $\mathbf{I}_2$ (**Figure 4)** which is important for discussion.

There is a significant fraction of long lived pick-off component due to *ortho* –positronium fraction: $\tau_3$ with its intensity $\mathbf{I_3}$ in this polymeric material and the temperature variation of the same is shown in **figure 5 and 6** . respectively.

As is noticeable from the positron life time results that the structural configuration of the giant polymeric sugar molecule is not a stable one at different temperature regimes but shows up repetitive pattern through-out the temperature region studied (270-360K), peaking at regular intervals, as shown by **Figure 4.** This part relates to the middle components $\tau_2$ and $I_{2\,\%}$ both. The analysis of the data shows this second positron lifetime component, in some cases is larger than the 'spin averaged' positronium lifetime(0.5ns). But it was not possible further to obtain a second longer life time component (i.e. a four life time component fit in the analysis was not possible). If we assume these middle components arise because of the crystalline part of the polymeric structure, then stacking pattern [19] of the fructose units with –O- bridged side chains certainly play a part. Simply, the strong overlap of positrons in their diffusion process[20] with the electron rich entities in the molecular chain(refer to **Figure 1**), namely vicinity of O atoms could be held responsible for higher decay rate, thereby lowering of the life time $\tau_2 = 1/\lambda_2$ and consequent increase in $I_2$. Alternatively, the stacking molecular space may not have a room for a clear cut longer positronium component but positrons sampling the area exhibit slightly higher values of $\tau_2$ for that domain with lesser annihilation probability. As a result we find a recurring periodicity in the lifetime values, in concurrence of the molecular structure. A simple diffusion model limited by trapping of positrons can be presented to explain the non monotonous behavior in the trend of $\tau_2$ values and derive the diffusion path length[21] through the short electron deficient apolar stacking domains interposed by electron rich regions due to polar H-O**:** groups on the polymeric structure of the molecule.

The diffusion path length has been calculated as: $\mathbf{L} = \sqrt{D\ 1/\lambda_{eeff}}$ ……………….(1)

where the positron diffusion coefficient D in the crystalline polymers is taken to be to $0.1 cm^2/sec$,[ 22] and $\lambda_{eff}$ has been calculated from : $\lambda_{eff} = 1/\tau_b + \kappa_r$ ………….. ……(2)

Here, $\tau_b$ is the bulk life time extracted from the positron data and $\kappa_r$ is the positron trapping rate as function of position[23].

By computing the difference of diffusion path lengths of positrons as a probe, at the different temperatures, starting from minima (corresponding to strong electric field of the polar group) to the maxima of the curve in **Figure 4**., nearly attains the height of the stacking distance (in terms of ~ 4 Å length) and from there back to minima completes one such unit in the polymeric structure. Thus the gap in between the two electron rich domain could be nearly the order of ≤ 1nm. This simple qualitative calculation corroborates with the 'd' lattice space data due to electron diffraction studies on inulin by reference 24 that is more prominently 8.35Å. Thus from the non-monotonous trend of the positron lifetime, $\tau_2$, one can assess the local organization of the oligomeric configuration of the sugar molecule, which would mean a higher annihilation rate from the higher electro negative site ( i.e. **:**O-H sites) with corresponding increase of intensity percentage, whereas a relaxed annihilation rate and lower intensity fraction from the electron deficient apolar region.

However, the long lived ortho positronium pick-off components ($\tau_3$) > 1ns arise with appreciable intensities because of the amorphous faction and also due to overall increase in the free volume [24] throughout the total molecular array, with rise in temperature and also owing to the surface structure (increased surface to volume ratio) of the nano grains. $\boldsymbol{\lambda_{pick\text{-}off}}$ is the reciprocal of annihilation rate due to long lived *ortho*-positronium fraction (a quasi stationary bound state of $e^-$ and $e^+$ with parallel spins residing in the free space available) which is described quantum mechanically as :

$$\lambda_{pick\text{-}off} = 4\pi r_0^2\, c\, d\, Z_{eff} \int d^3r\, |\psi_{ps}(r)|^2 \quad \text{.......................(3)}$$

where, $\int d^3r\, |\psi_{ps}(r)|^2$ is the overlap integral of positronium wave function in that local environment, with the number density of molecule 'd' and $Z_{eff}$ taken as number of singlet electron available per molecule, $r_0$ is the classical electron radius and c is the velocity of light.

The free volume radii could be calculated, wherein the size of the same can be related to *ortho*-positronium pick-off lifetime, using a simplified quantum mechanical model [25], that describes Positronium being trapped in tiny spherical cavities (free space) under an attractive potential, with diffused boundaries of the cavity walls, with their radii taken as R+ δR (δR carrying all the corrections due to roughness of the walls and has the length parameter, where annihilation with electrons are likely). The life time versus the resulting values of the cavity radius R were tested and found to lie very closely on a straight line and with an extremely simple relationship as follows:

$$\tau_{pick\text{-}off} = 1.88R - 5.07 \quad \text{..........(4)}$$

The comparison of this diffused boundary model with that of the popularly used Tao-Eldrup model [26,27] has long been exemplified [28]. Although the former appears pretty simple yet it assumes a lot of corrections due to finite size of positronium, diffusivity of the walls and curvature dependence of surface tension in case of liquids [29] etc., all lumped into δR for simplicity, but it does not appear as an adjustable free parameter as is in the case of Tao –Eldrup formulation.

In **figure 5**, the pick-off life time ($\tau_3$), the free volume radii also show the similar variance as they are related. It is noteworthy that the central thermotropic transition point in the molecular structure lies in the region ~320-330K . However, the smaller segments in the giant molecular structure may also contribute to a certain extent in a complex way, in the molecular packing and free volume at the given temperature regime, as the average life time or the pick –off life time of positronium fraction maps a non-monotonous electron density variation, before and after the transition zone. The free volume radii increase with increase in temperature thereafter and could be due to loss in structure because of the glass transition temperature of the polymer[30]. Additionally, the fraction of positrons that annihilate as pick-off component of *ortho*-Ps, the o-Ps intensity ($I_3$), (**Figure 6a**) is related to the formation probability of positronium which may be proportional to the free volume fraction/ number of free volume elements in a polymeric compound provided there is no other chemical interaction of Ps with the substrate (like inhibition or oxidation etc). But the $I_3$% values show a reverse trend, and the effect of the same is carried over in the average long lived life time component (**figure 6 b**), meaning thereby that some inhibition effect may decrease ortho-positronium formation probability.

**b) Doppler Broadening of Annihilation Radiation (DB) study** :

**Figure 7**. shows the line shape parameters ('S' and 'S' vs 'W') for the DB spectrum of positron annihilation radiation[21] of Inulin sample when subjected to different temperature regimes from 270-360K. The 'S' parameter represents the fraction of low/narrow momentum component of the annihilating electron positron pairs and shows a non monotonous behavior denoting the variation of the free volume fraction with temperature. Positrons/positronium trapped in the free volume region, may result in lowered pair momentum. This actually affirms that the molecular configuration is not very stable, either a part (or a particular molecular segment) in the polymeric molecule could be responsible for such a transition which also changes the overall free volume of the system. The results apparently appear to follow a periodic pattern, which could result from a stacking pattern of fructose units. 'W" parameter arises due to the high momentum components

of the annihilation radiation in the polymeric structure, and the 'S vs 'W' shows a linear variation, possibly since only one type of free space (similar free volume structure) as positron trap is available but an island of the points crowding around a regime as shown by the arrow. This correspond to a temperature of transition region( 313-318 K). However, from the usage point of view of such a giant sugar molecule and from these results, one can be certain that above the freezing temperatures to the room temperature there is a transition of the polymeric sugar structure, and then the major transition occurs around $\geq$ 320K from the temperature region investigated.

Further it is deemed necessary to mention that the positron parameters reported here are irreversible after the temperature treatment, and the substance is visibly transformed to a different physical state as observed from the sample vial.

**c) DSC and thermal analysis:**

In quest of obtaining additional insight into the nature of the phase transition of inulin as regards the structural change revealed by calorimetric measurements [9], a systematic study with inulin sample (which is stored under vacuum desiccation) was performed. In our study it is observed (Figure 8a) that the sample undergoes configuration/structural phase transition showing a strong endothermic peak during heating, which is irreversible, as is evident from the cooling cycle. The onset of peak in the transition occurs at 319 K (T) whereas the maximum occurs at 324K ($T_P$), respectively. Although the thermodynamic transition is noticed at the temperature 319 K, it goes on over a broad region till 400K.The enthalpy change ($\Delta H$)associated with the transition, obtained from this data after averaging the results from three different scans is $\Delta H$ ($T_{\text{p}}$) ~379 J/gm.

The thermo-gravimetric analysis shows that the sample is stable up to 319 K, shown in **Figure 8b.** A small mass loss takes place between the temperatures 319 - 383K. It is clear from thermal plot that the mass loss up to383K is 7.5% which could correspond to degradation of H--OH from the long chain of the fructose units mainly from within the sample and also due to inherent moisture loss( since not perfectly "dry").

## 4. CONCLUSION:

Inulin is a nanometer size carbohydrate polymer with a partial crystalline symmetry, having a hydrogen bonded supra molecular structure and is liable to thermotropic phase transition. Positron annihilation studies have probed the molecular configuration changes as well as and free volume changes with temperature. This structural transition beyond 330 K has arisen due to the glass transition in inulin. Thermotropic phase transition has been affirmed through an independent thermal analysis. Detailed studies on inulin in its solid and solution states are required by positron annihilation spectroscopy for further understanding of the properties of the stacked sugar (fructose) units in its structure.

## 5. ACKNOWLEDGEMENTS

The first author BNG acknowledges Experimental Condensed Matter Physics Division, SINP for the technical help received for X-ray Diffraction data and Applied Material Science Division, SINP for IR spectroscopic data. Soma Roy is acknowledged for all kinds of technical help.

CAPTIONS TO THE FIGURES

1. Schematic representation of the inulin molecule showing the fructose units joined by a β(2→1) glycosidic bond.

2. X-ray diffraction pattern of inulin powder (stored under vacuum desiccation ) showing a strong peak at respective Bragg angles. The dotted line shows the subtracted portion due to amorphous scattering fraction.

3. FT –IR spectrum for the assay of the functional groups of the inulin sugar units.

4. The middle component of the life time spectrum a) $\tau_2$ and b) $I_2$ , in the inulin sample showing the periodic ripple with varying temperature.

5. Pick-off component of the ortho -positronium fraction, along with the temperature variation a) life time and b) the free volume radius , in the inulin sample.

6. The line shape parameters from the D.B. spectrum showing a) 'S' parameter with their non monotonous behaviour upon temperature variation, for inulin sample.
   b) 'S' vs 'W' parameter showing a linear variation.

8. The thermal analysis of the sample a) DSC: scan showing the thermotropic transition, and b) TGA: graph exhibiting the loss of mass from the molecule due to heating.

Figure 1. : INULIN

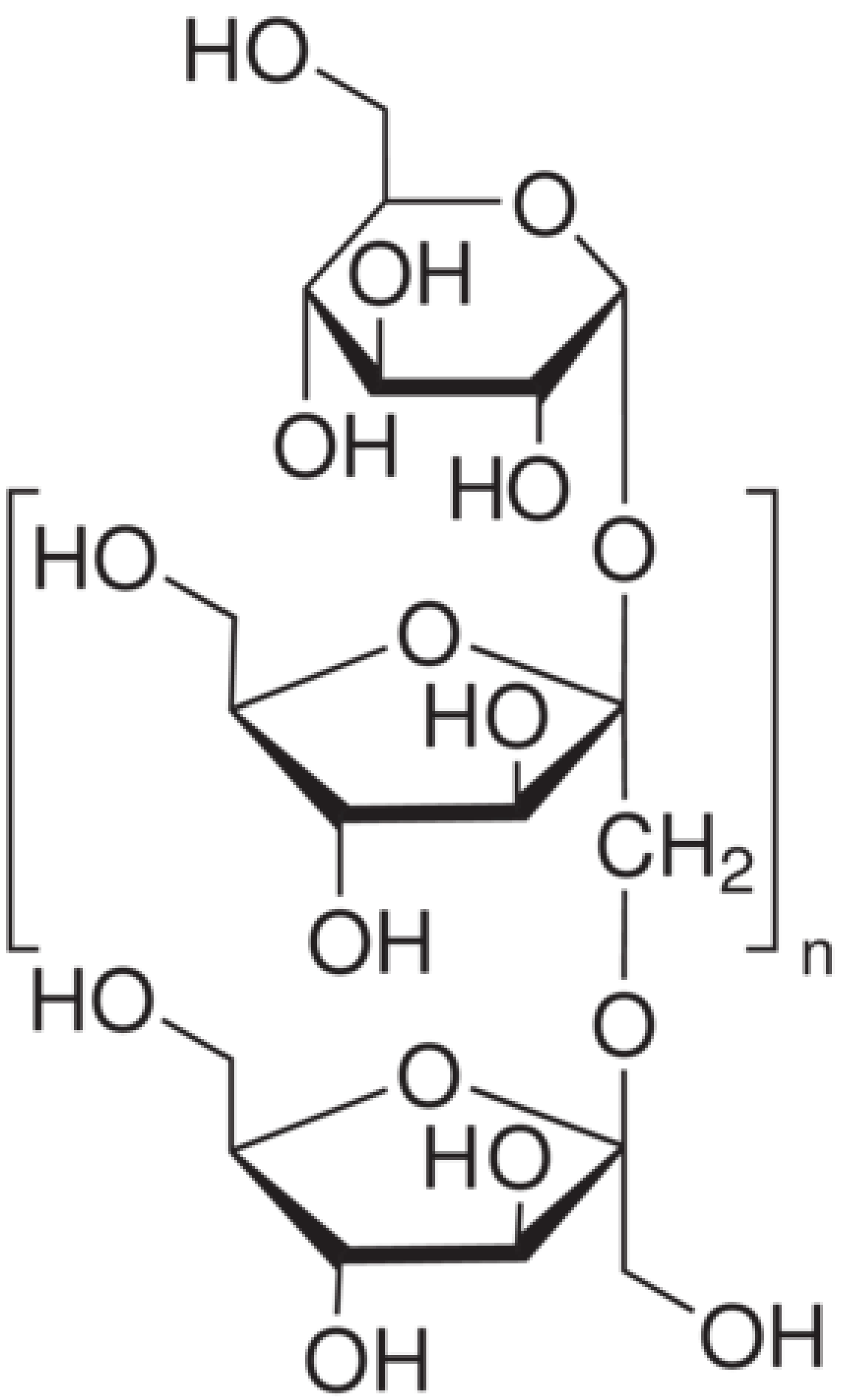

Scheme 1. Structure of Inulin

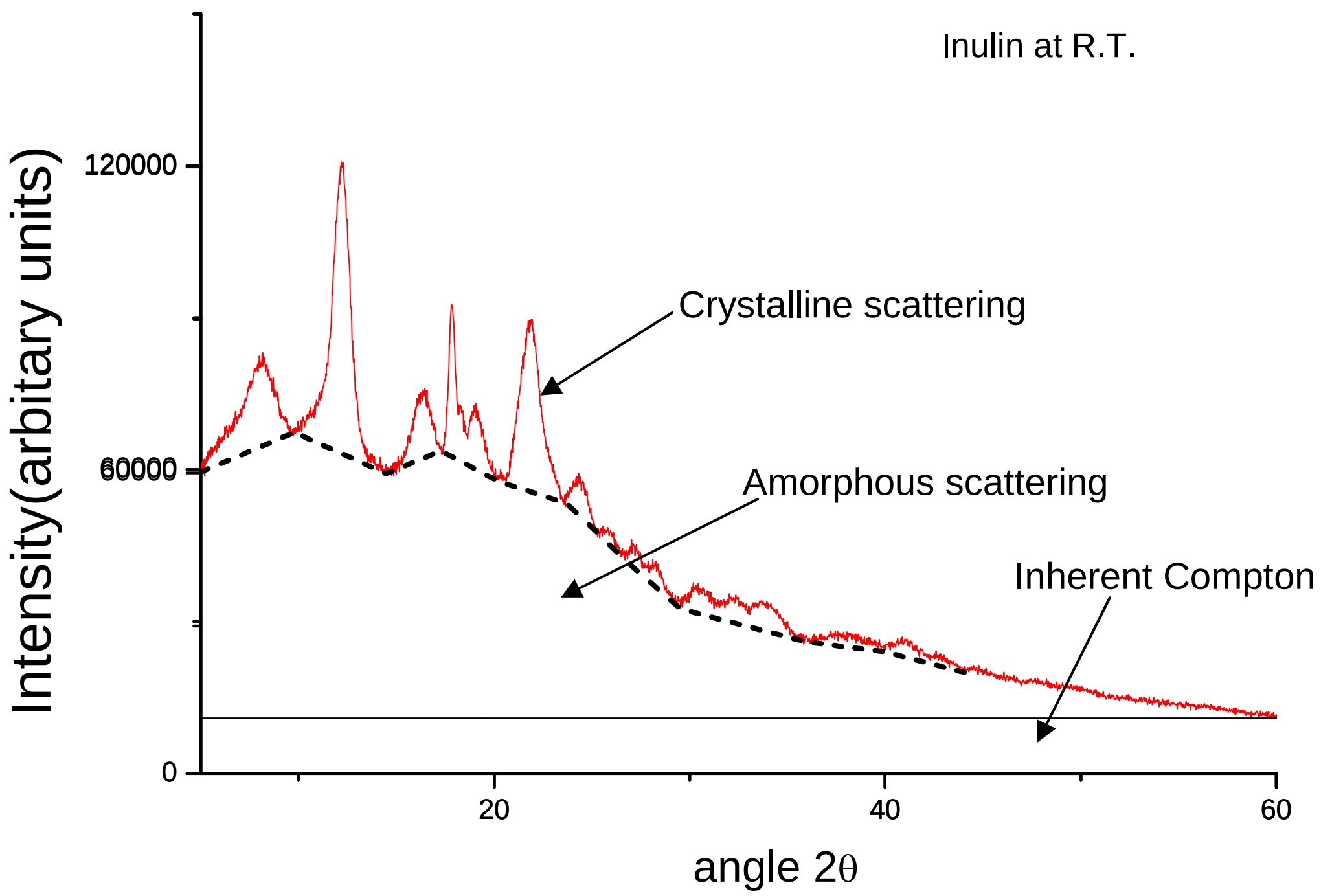

Figure 2. X-ray Diffraction spectrum

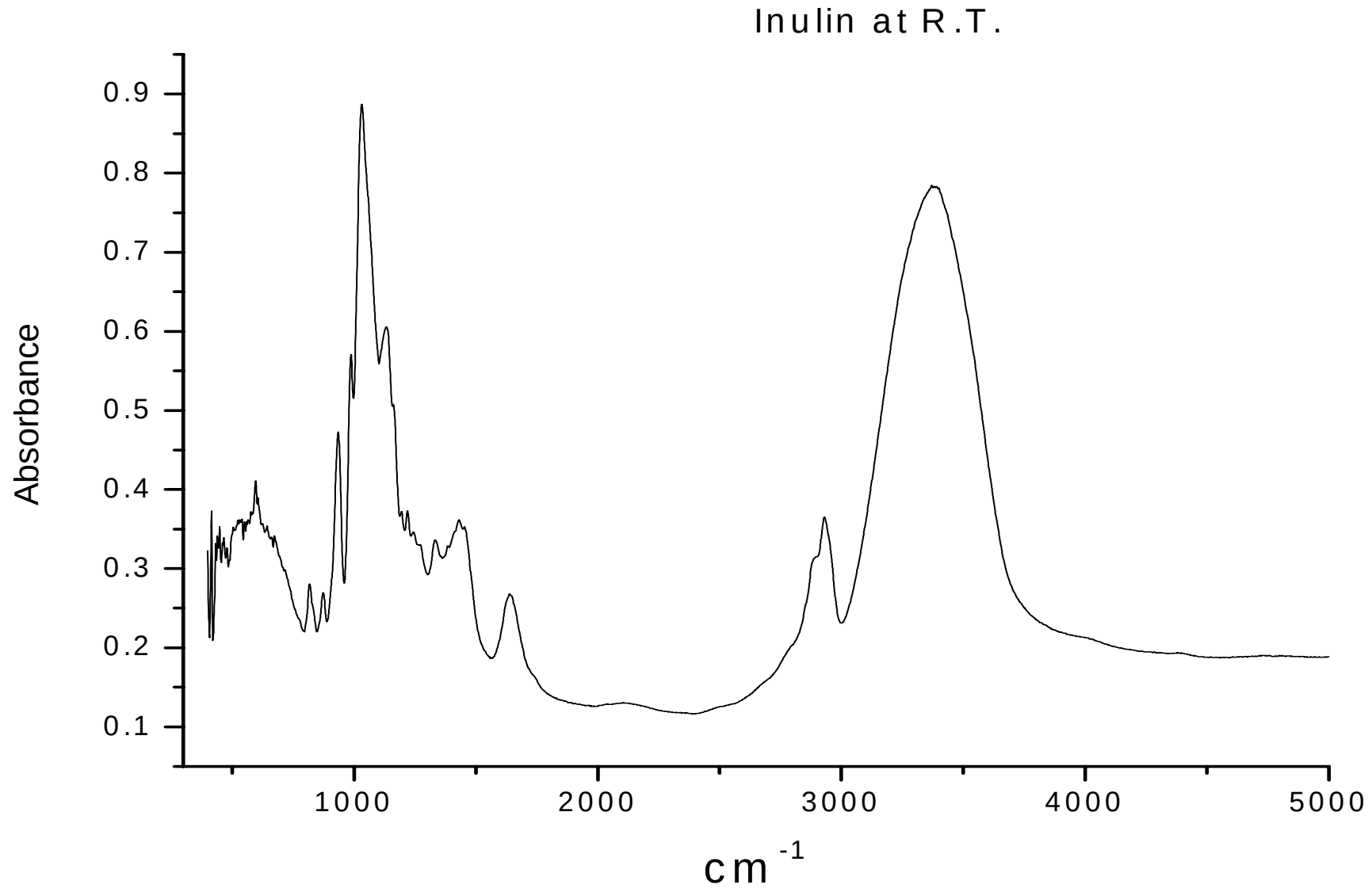

Figure 3. IR Spetrum of Inulin

Figure 4a,b

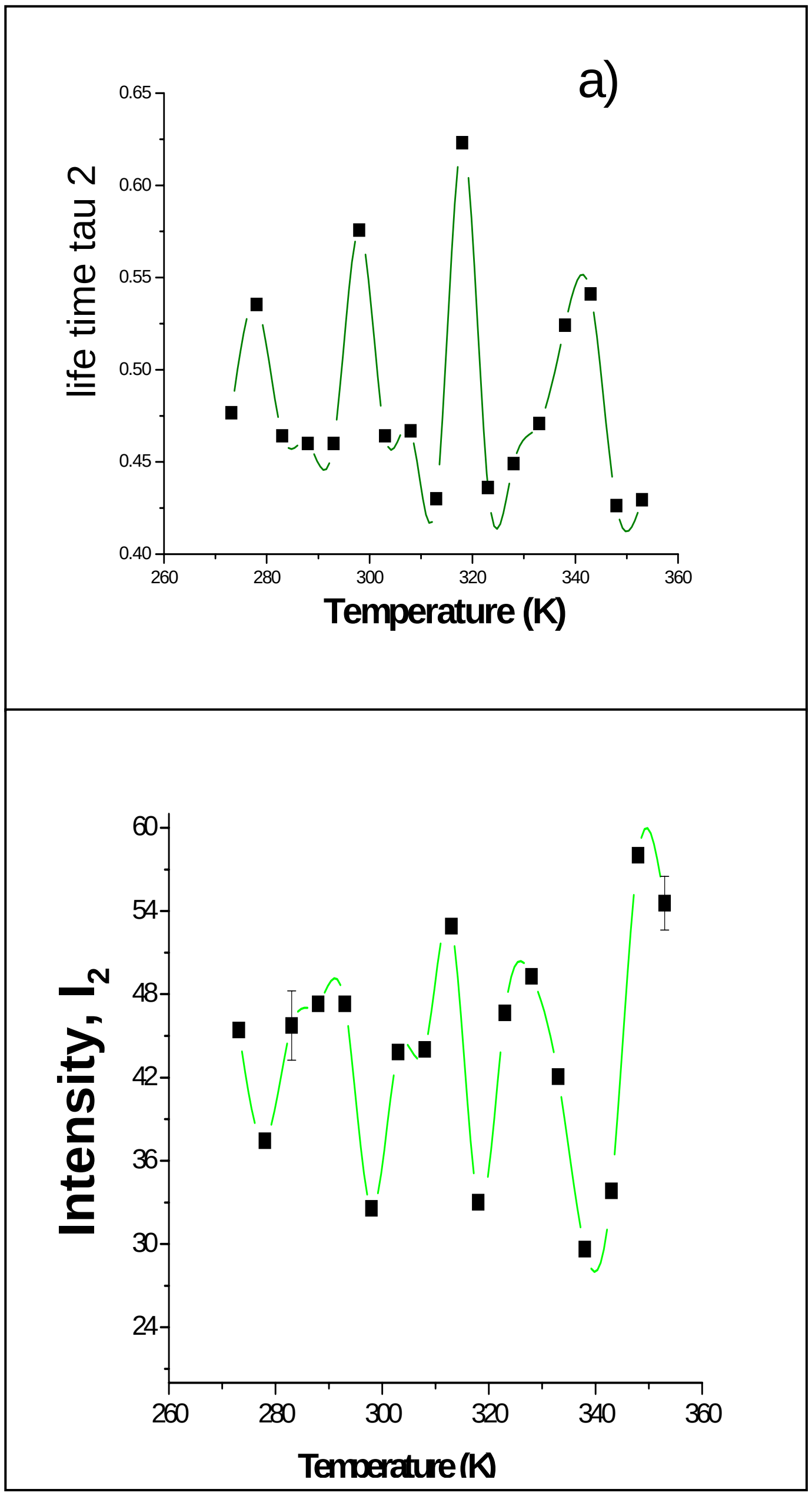

Figure 5 a,b,

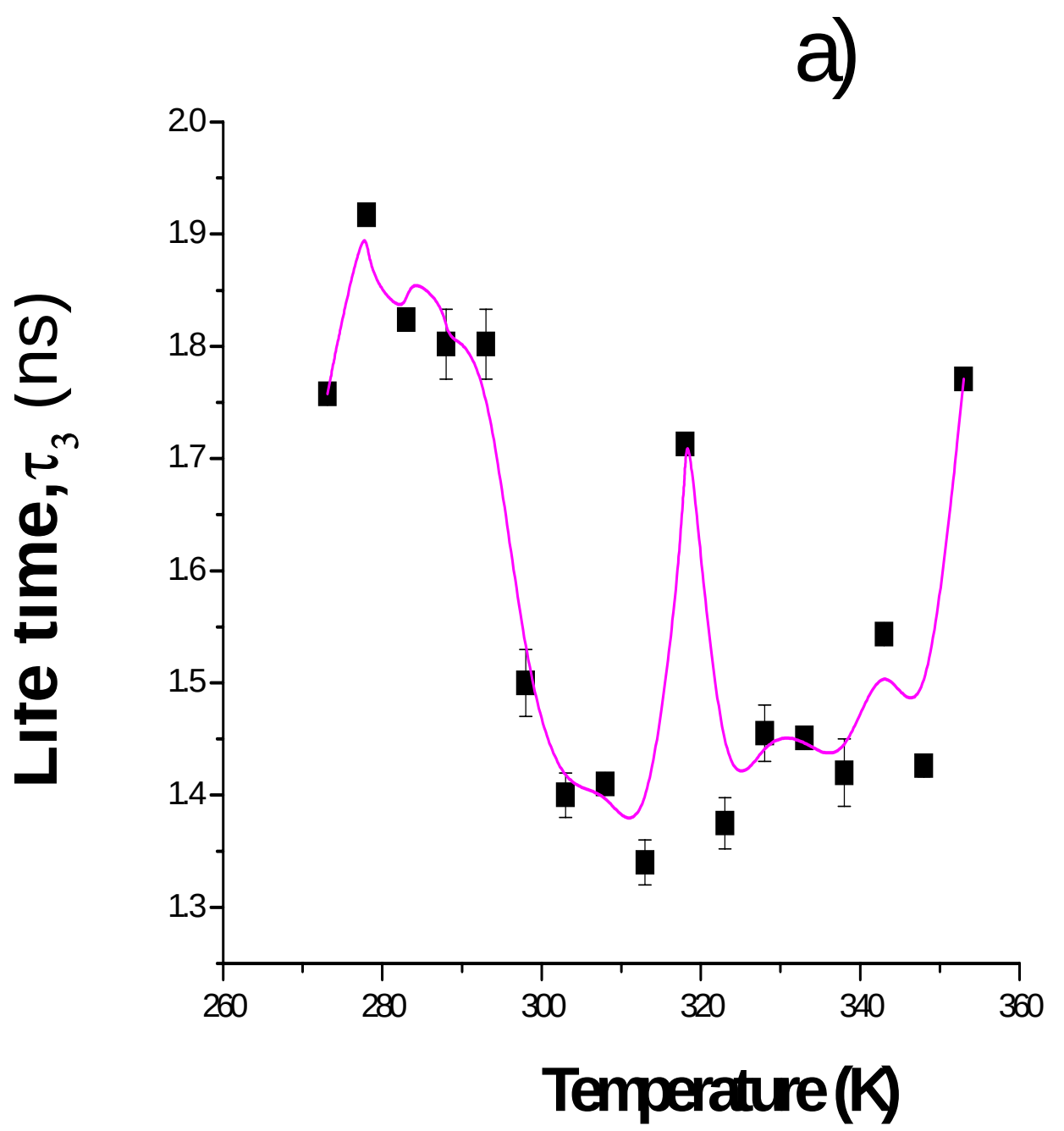

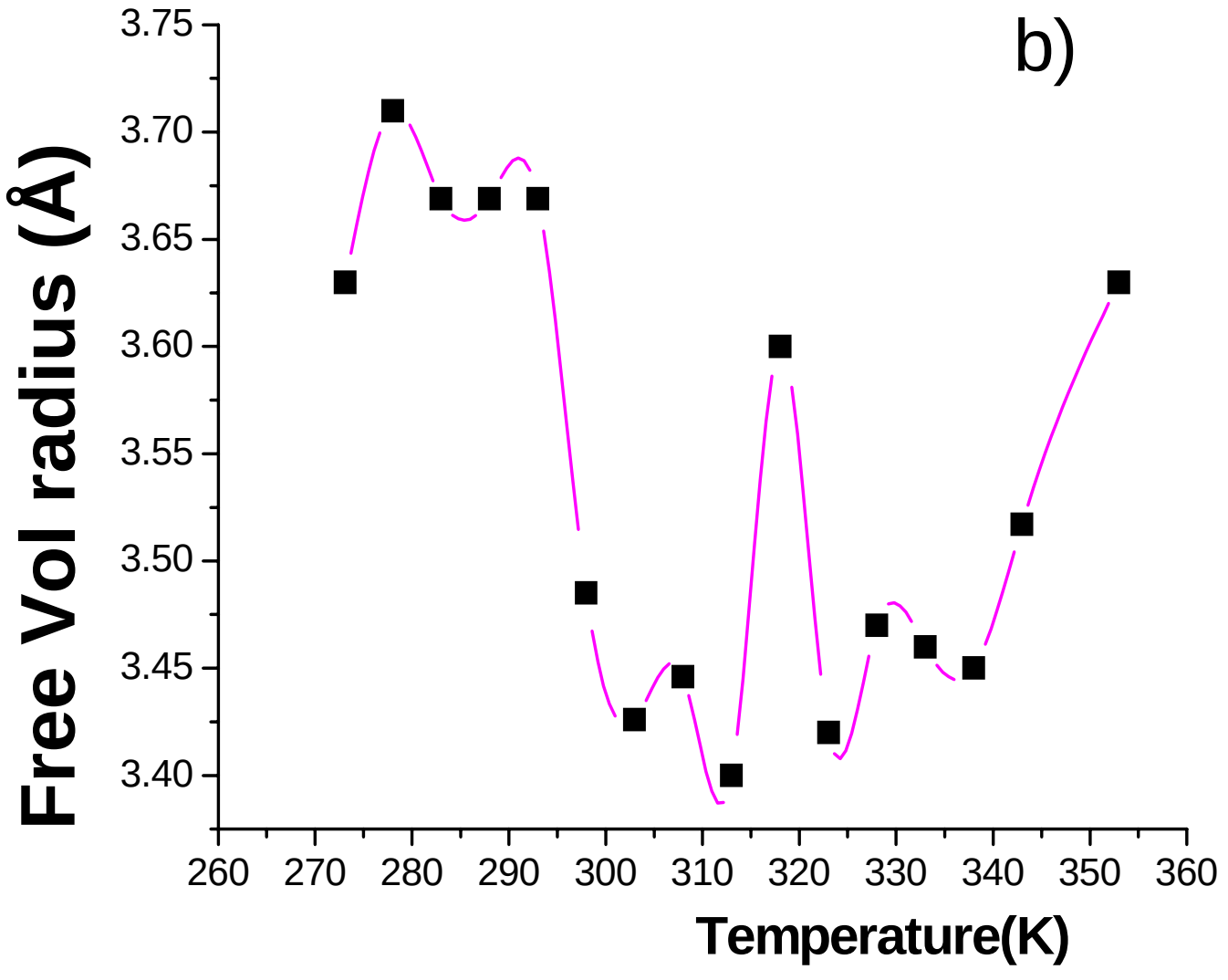

**Figure 6. a,b.**

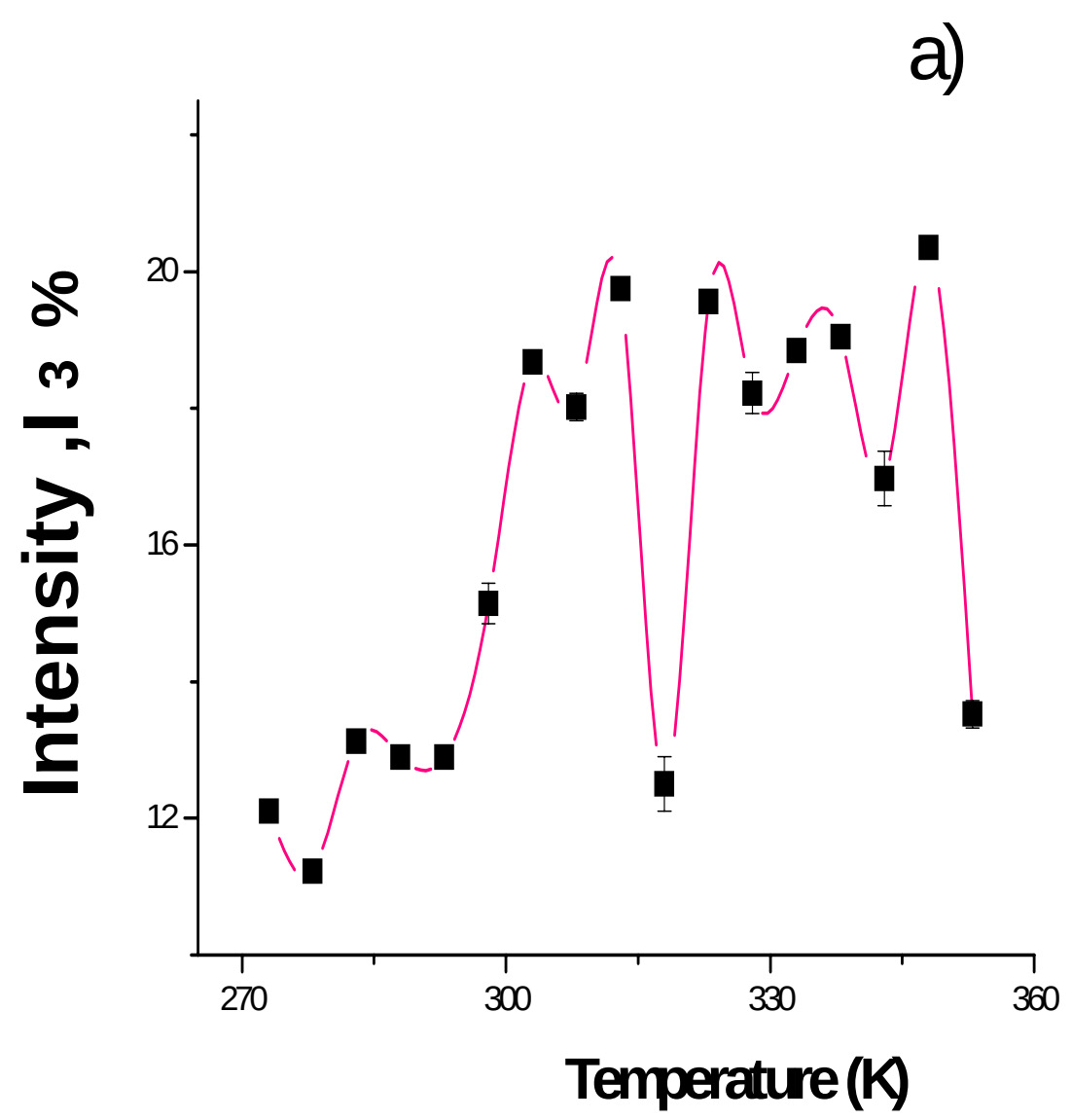

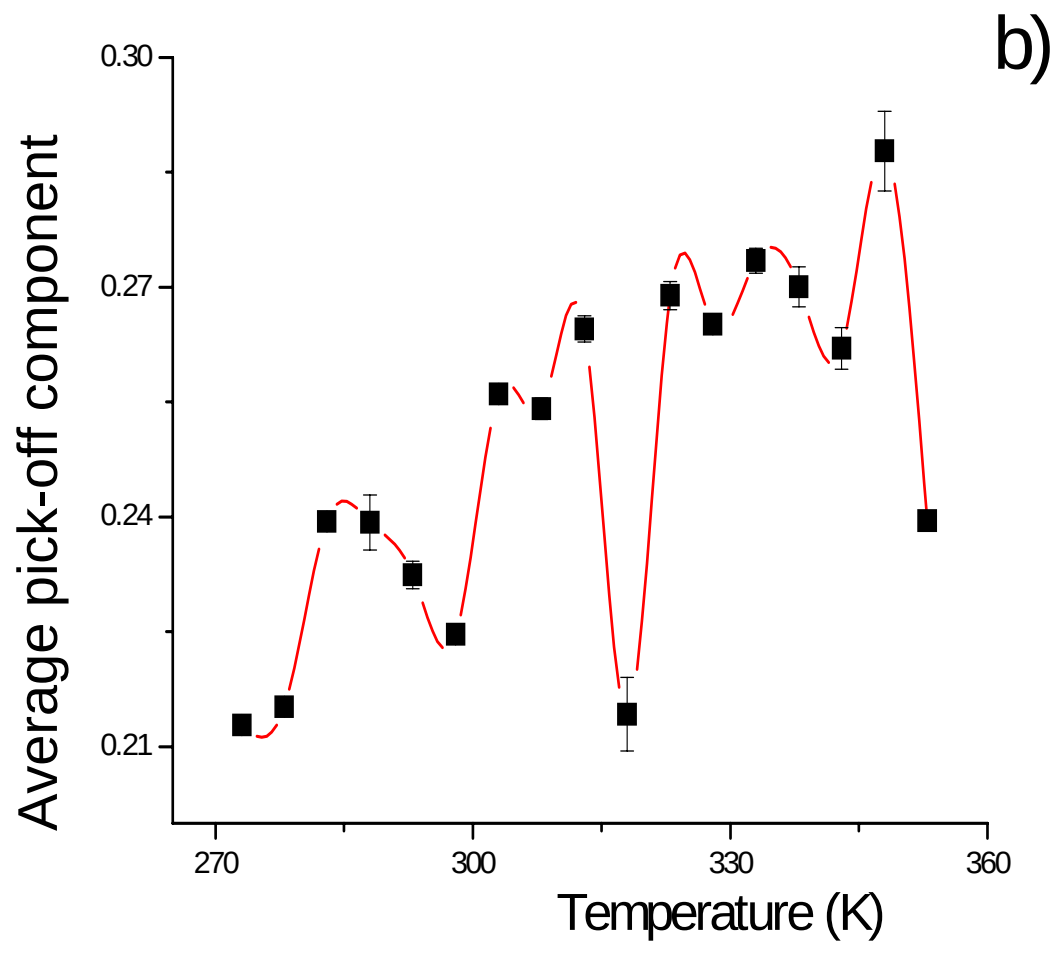

Figure 7. a, and b

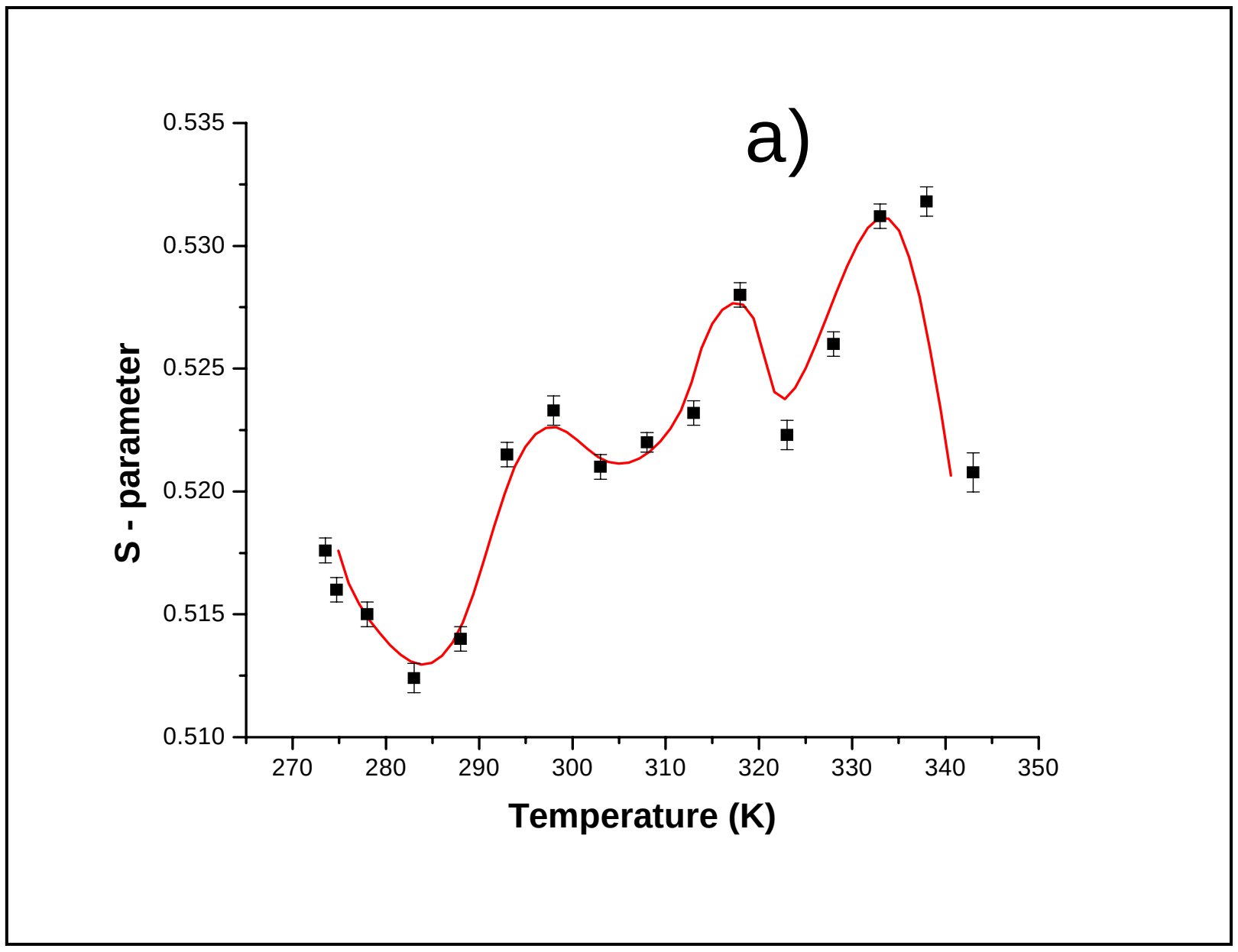

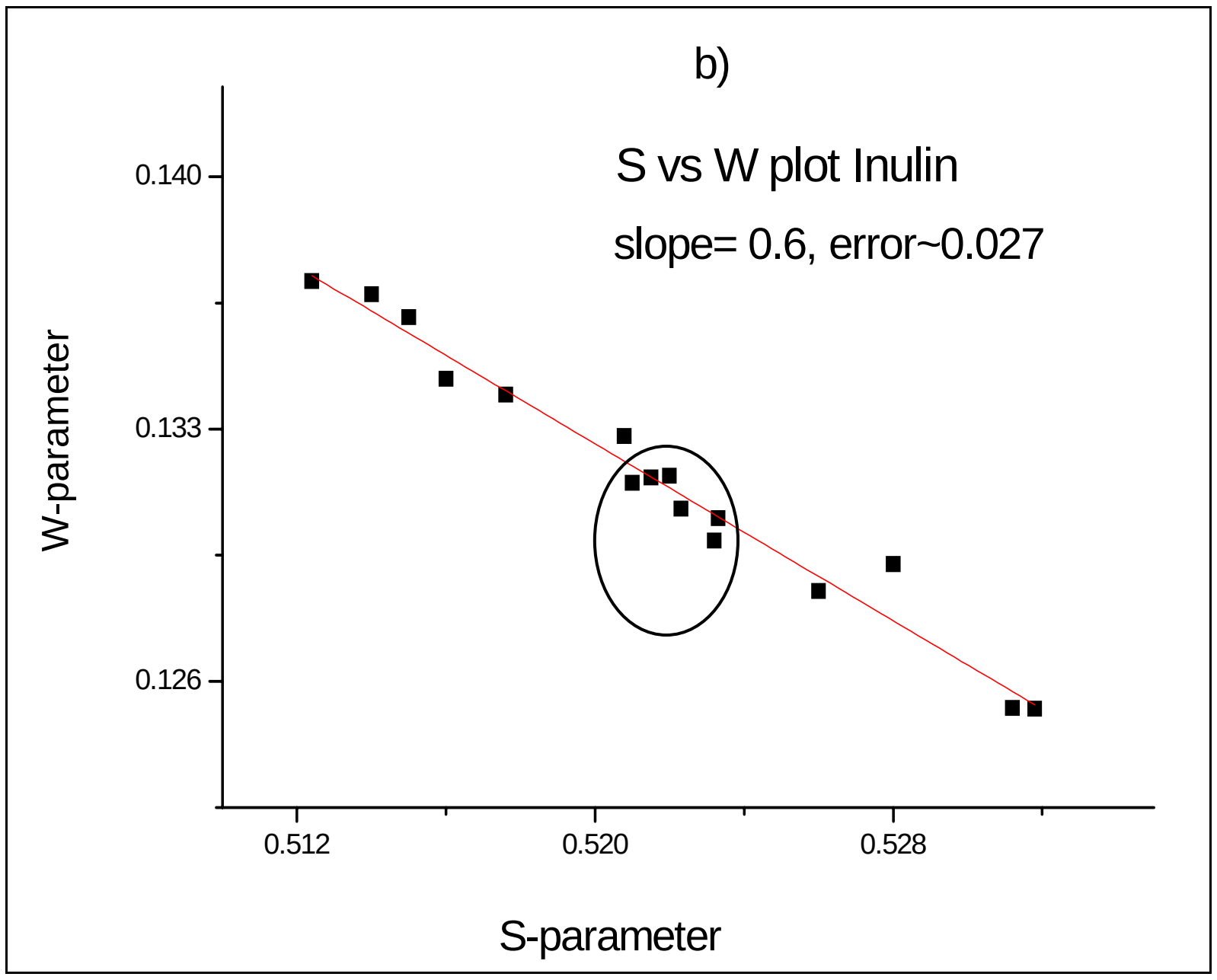

Figure 8 a, b

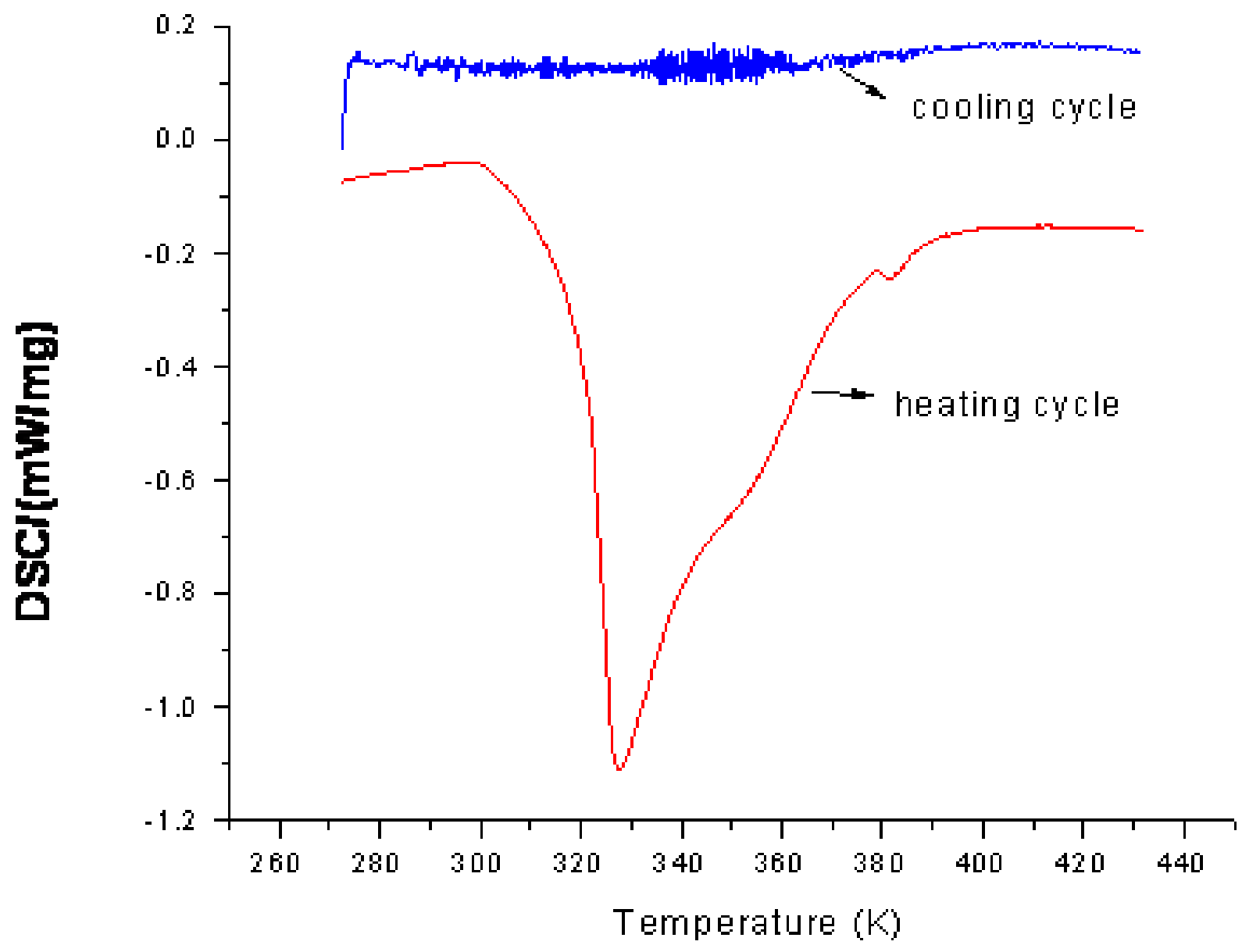

Fig.8a

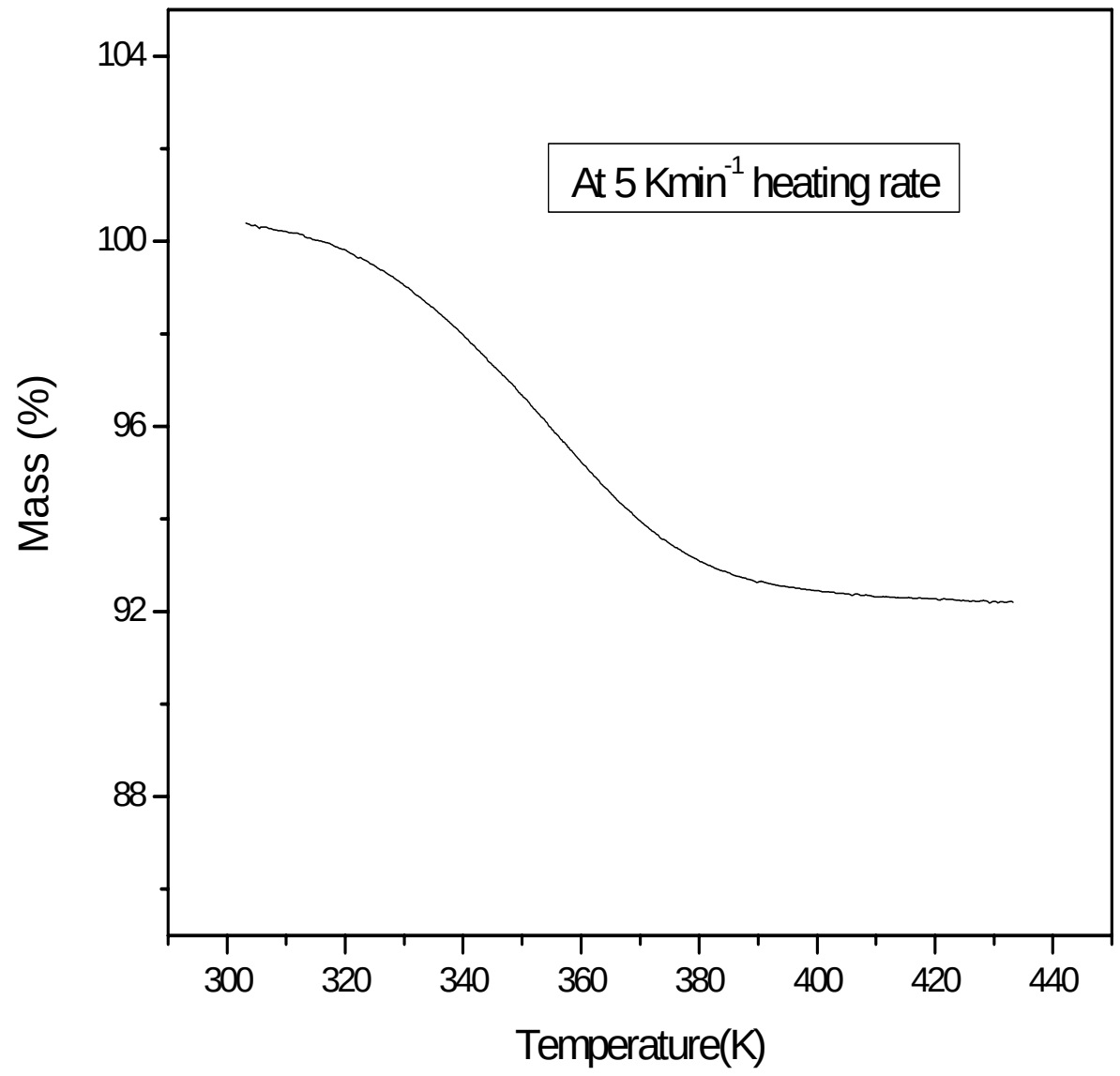

Fig 8b